\documentclass[submission,copyright,creativecommons]{eptcs}
\providecommand{\event}{HCVS 2021} 
\usepackage{breakurl}             
\usepackage{underscore}           
\usepackage{amsthm}
\usepackage{graphicx}
\usepackage{cite}
\usepackage{color}

\newcommand\DEF{\mathbin{:=}}
\newcommand\seq[1]{\tilde{#1}}
\newcommand\COL{\mathbin{:}}
\newcommand\HFLZ{\textrm{HFL(Z)}}

\newcommand\nuHFLZ{$\nu$\textrm{HFL(Z)}}
\newcommand\nuHFL{$\nu$\textrm{HFL}}
\newcommand\form{\varphi}
\newcommand\Some[1]{\langle #1\rangle}
\newcommand\All[1]{[#1]}
\newcommand\INT{\mathtt{int}}
\newcommand\Prop{\mathtt{o}}
\newcommand\sty{\tau}
\newcommand\fty{\kappa}
\newcommand\FALSE{\mathtt{false}}
\newcommand\TRUE{\mathtt{true}}
\newcommand\Z{\mathbf{Z}}
\newcommand\mult{\mathtt{mult}}

\newcommand\M{\mathcal{M}}
\newcommand\order{\mathtt{ord}}
\newcommand\imp{\Longrightarrow}
\newcommand\dual[1]{\overline{#1}}
\newcommand\To{\Rightarrow}

\title{An Overview of the HFL Model Checking Project}
\author{Naoki Kobayashi
\institute{The University of Tokyo\\ Tokyo, Japan}
\email{koba@is.s.u-tokyo.ac.jp}
}
\def\titlerunning{HFL Model Checking Project}
\def\authorrunning{N. Kobayashi}
\theoremstyle{definition}
\newtheorem{example}{Example}
\begin{document}
\maketitle

\begin{abstract}
  In this article, we give an overview of our project on
  higher-order program verification based on HFL (higher-order fixpoint logic)
  model checking.
  After a brief introduction to HFL, we explain how it can be applied to program
  verification, and summarize the current status of the project.
\end{abstract}

\section{Introduction}
In this article, we give a brief overview of our project on automated
verification of higher-order programs based on
(a variation of) Viswanathan and Viswanathan's higher-order fixpoint logic
(HFL)~\cite{Viswanathan04}.
HFL is a higher-order extension of the modal $\mu$-calculus and is
strictly more expressive than the modal $\mu$-calculus, but
the HFL model checking problem for finite state systems remains decidable.

In Section~\ref{sec:hfl}, we first review HFL, and \HFLZ{}, an extension of HFL with integer arithmetic,
and show that \HFLZ{} may also be viewed as an extension of Constrained Horn Clauses
(CHC)~\cite{DBLP:journals/sttt/DelzannoP01,DBLP:conf/vmcai/JaffarSV06,DBLP:conf/sas/PeraltaGS98,Bjorner15} with higher-order predicates and fixpoint alternations.

In Section~\ref{sec:red},
we show how various program verification problems can naturally be reduced to
\HFLZ{} model/validity checking problems.
Our program verification framework based on \HFLZ{} can be considered
a generalization of CHC-based program verification framework~\cite{DBLP:journals/sttt/DelzannoP01,DBLP:conf/vmcai/JaffarSV06,DBLP:conf/sas/PeraltaGS98,Bjorner15}.
In Section~\ref{sec:solve}, we summarize our methods for automatically solving
the \HFLZ{} model/validity checking problems, using higher-order model checkers and
CHC solvers as backends.

This article is intended to be a non-exhaustive survey of
HFL-based approaches to program verification.
The main objective is to provide references to technical papers and clarify
how they are connected with each other; 
the explanation of each topic is admittedly short and cryptic.
\section{Higher-Order Fixpoint Logic and its Relationship with CHC}
\label{sec:hfl}

\subsection{Higher-Order Fixpoint Logic}
\label{subsec:HFL}
The syntax of \HFLZ{} formulas and types is given as follows.
\[
\begin{array}{l}
  \left.\qquad\qquad
  \begin{array}{l}
    \form \mbox{ (formulas) } ::=
    x^\sty \mid \form_1\lor \form_2 \mid \form_1\land\form_2\\
\qquad\qquad  \qquad  \mid  \Some{a}{\form}\mid \All{a}{\form} \qquad\ \mbox{ (modal operators) }\\
\qquad\qquad  \qquad     \mid \mu x^\fty.\form \mid \nu x^\fty.\form \quad\mbox{ (fixpoint operators) }\\
\qquad\qquad  \qquad     \mid \form_1\form_2 \mid \lambda x^\sty.\form\quad\ \ \mbox{ (\(\lambda\)-abstractions and applications)}\\
    \end{array}
  \right\} \mbox{pure HFL}\\
  \qquad\qquad\qquad\qquad  \ \ \qquad \mid \form\, e \mid e_1\le e_2\qquad
  \mbox{ (extension with integers)}\\
  \ \   e \mbox{ (integer expressions) }::= n \mid x^{\INT} \mid  e_1+e_2\\
  \qquad\qquad\qquad \sty \mbox{ (types) }::= \INT \mid \fty\\
  \qquad \ \fty \mbox{ (predicate types) }::= \Prop \mid \sty \to \fty\\
\end{array}
\]

Here, 
each variable (denoted by the metavariable \(x, y, \ldots\)) has its own type
(specified as a superscript), and we consider only
well-typed formulas and expressions; for example,
in \(\form_1\lor \form_2\), both \(\form_1\) and \(\form_2\) must have
the type \(\Prop\) of propositions.
We often omit the type annotation of a variable.
Each formula must have a predicate type, and integer expression must
have type \(\INT\). See \cite{KTW18ESOP} for typing rules.

The first three lines of the definition of formulas correspond to the
syntax of the modal \(\mu\)-calculus~\cite{Kozen83,Automata} in negation normal form
(if the types \(\fty\) and \(\sty\) are restricted to the type \(\Prop\)
of propositions), and the first four lines
correspond to the syntax of pure HFL~\cite{Viswanathan04}.
Each formula of type \(\Prop\) describes a property
of labeled transition systems. The formula \(\Some{a}\form\) means
that there exists a transition labeled \(a\) after which the proposition \(\form\)
holds, while the formula \(\All{a}\form\) means that, after any transition
labeled \(a\), the proposition \(\form\) holds.
The formulas \(\mu x^\fty.\form\) and
\(\nu x^\fty.\form\) respectively denote the least and greatest predicates \(x\)
such that \(x=\form\).
For example, \(\mu x^\Prop.x^\Prop\) and \(\nu x^\Prop.x^\Prop\)
respectively denote the least and greatest propositions such that \(x=x\),
i.e., \(\FALSE\) and \(\TRUE\) respectively; henceforth, we treat
\(\FALSE\) and \(\TRUE\) as propositional constants.
The fifth line is for the extension with integers. We have only
constants, additions, and inequality constraints on integers, but
other predicates and operations, as well as quantifiers, are definable,
as shown below.

\begin{example}
  \label{ex:quantifiers}
  Let \(\form\) be a formula of type \(\INT\to \Prop\), and
  \(\psi\) be \(\nu x.\lambda n.\form(n)\lor x(n+1)\).
  Then we have
  \[
  \begin{array}{lcll}
    \psi(0) &\equiv& (\lambda n.\form(n)\lor \psi(n+1))0  &\mbox{ (by unfolding $\psi$)}\\
     &\equiv& \form(0)\lor \psi(1)  &\mbox{ (by $\beta$-reduction)}\\
     &\equiv& \form(0)\lor (\lambda n.\form(n)\lor \psi(n+1))(1)  
     &\mbox{ (by unfolding $\psi$)}\\
     &\equiv& \form(0)\lor \form(1) \lor (\lambda n.\form(n)\lor \psi(n+1))(2)  
     &\mbox{ (by $\beta$-reduction)}\\
     &\equiv& \form(0)\lor \form(1) \lor \form(2) \lor \cdots.
  \end{array}
  \]
  Thus, \(\psi(0)\) denotes \(\exists x\ge 0.\form(x)\).
  Similarly, \(\forall x\ge 0.\form(x)\) can be expressed by
  \((\nu x.\lambda n.\form(n)\land x(n+1))0\),
  and \(\forall x\in \Z.\form(x)\) (where \(\Z\) is the set of integers)
  can be expressed by
  \((\nu x.\lambda n.\form(n)\land x(n-1)\land x(n+1))0\).
  The multiplication can be expressed as a ternary predicate
  \(\mult(x,y,z)\) (which means \(x\times y=z\)), e.g., by:
  \[\mult \DEF \mu u.\lambda (x,y,z).y=z=0 \lor (1\le y\land u(x, y-1,z-x))\lor
  (y+1\le 0\land u(x, y+1,z+x)).\]
  Here we have used tuple notations and subtractions (\(-\)) 
  for readability; as usual,
  subtractions can be defined by using additions and existential
  quantifiers.
 \qed
\end{example}

As shown in the example below, pure HFL is already strictly more expressive
than the modal $\mu$-calculus.
\begin{example}
  \label{ex:hfl}
  Consider \(\form\) be the formula \(\nu x.\lambda y. y\lor \Some{a}x(\Some{b}y)\).
  Then, we have:
  \[
  \begin{array}{lcll}
    \form(\Some{c}\TRUE) &\equiv&
    (\lambda y. y\lor \Some{a}\form(\Some{b}y))(\Some{c}\TRUE)
     & \mbox{ (unfolding of $\form$)}\\
   &\equiv& \Some{c}\TRUE\lor \Some{a}\form(\Some{b}\Some{c}\TRUE)
     & \mbox{ ($\beta$-reduction)}\\
    &\equiv& \Some{c}\TRUE\lor \Some{a}
    (\Some{b}\Some{c}\TRUE\lor \Some{a}\Some{a}\form(\Some{b}\Some{b}\Some{c}\TRUE))
     & \mbox{ (unfolding, followed by $\beta$)}\\
    &\equiv& \Some{c}\TRUE\lor \Some{a}\Some{b}\Some{c}\TRUE
    \lor \Some{a}^2\Some{b}^2\Some{c}\TRUE\lor \cdots
  \end{array}
  \]
  Thus, \(\form(\Some{c}\TRUE)\) means that there exists
  a transition sequence of the form \(a^nb^n\) after which
  a \(c\)-transition is enabled.  \qed
\end{example}

We write \(\M\models \form\) when a labeled transition system \(\M\) satisfies
\(\form\). We omit the formal semantics of \HFLZ{}~\cite{KTW18ESOP}.
The \emph{model checking problem} for \HFLZ{} is
the problem of checking whether \(\M\models \form\) holds,
given a finite labeled transition system \(\M\) and
a \HFLZ{} formula \(\form\) of type \(\Prop\).
The \emph{validity checking problem} is a special case:
it is the problem of checking whether \(\M_0\models \form\) holds,
where \(\form\) is a \HFLZ{} formula
of type \(\Prop\) without modal operators (\(\Some{a}\), \(\All{a}\)),
and \(\M_0\) is a trivial model consisting of a single state without transitions.
We often just write \(\models \form\) for \(\M_0\models \form\) and say
``\(\form\) is valid'' when \(\models \form\) holds.
For pure HFL, the model checking problem (hence also the validity checking problem)
is decidable~\cite{Viswanathan04} and \(k\)-EXPTIME complete for
the order-\(k\) fragment~\cite{DBLP:journals/lmcs/AxelssonLS07}.
Here, the order of the model checking problem
is defined as the the largest order of types that occur
in the HFL formula, and the order of a type \(\sty\), written \(\order(\sty)\),
is defined by:
\[
\order(\Prop)=\order(\INT)=0
\qquad
\order(\sty\to\fty) = \max(\order(\sty)+1, \order(\fty)).
\]
For example, the order of the formula in Example~\ref{ex:hfl}
is \(1\).
For \HFLZ{}, both the model and validity checking problems
are undecidable due to G\"{o}del's incompleteness theorem (recall that quantifiers
and multiplications can be expressed, as discussed in Example~\ref{ex:quantifiers}).

\subsection{Relationship with CHC}
We now explain the connection between the \HFLZ{} validity checking problem
and the CHC satisfiability problem, through examples.
A little formal discussion on the correspondence between \HFLZ{} and CHC is found
in \cite{DBLP:conf/sas/0001NIU19}.
We assume that the reader is familiar with
Constrained Horn Clauses (CHCs); those not familiar with CHCs may wish
to consult~\cite{Bjorner15}, which is a good survey CHCs and their applications
to program verification.

Let us consider the following system \(S\) of CHCs.
\[
\begin{array}{rcl}
  \forall x, y, r.\; y=0\land r=0 &\imp& \mult(x, y, r)\\
  \forall x, y, r, s.\; y\ne 0\land \mult(x, y-1, s)\land r=s+x &\imp& \mult(x,y,r)\\
  \forall x, y, r.\; \mult(x,y,r) \land x> 0   &\imp& r\ge y.
\end{array}  
\]
The satisfiability of CHCs above (i.e., the existence of
an assignment of a predicate to the predicate variable \(\mult\))
is equivalent to the safety property of the following OCaml program
(that assertion failures never occur):
\begin{quote}
\begin{verbatim}
let rec mult(x,y) =
  if y=0 then 0
  else let s = mult(x, y-1) in s+x
let main x y = if x>0 then assert(mult(x,y)>=y)
\end{verbatim}
\end{quote}
Here, the ternary predicate \(\mult(x,y,r)\) in the CHCs intuitively
means that the return value of \(\mult(x,y)\) in the program is \(r\).

Let us now convert the satisfiability problem for the CHCs above to
the validity checking problem for a \HFLZ{} formula.
First, note that the first two clauses are equivalent to:
\[
\forall x,y,r.
(y=0\land r=0) \lor\exists s.(y\ne 0\land \mult(x,y-1,s)\land r=s+x)
\imp \mult(x,y,r).
\]
The least predicate \(\form\) that satisfies the condition above
is expressed by:
\[
\mu u.\lambda (x,y,r).(y=0\land r=0) \lor\exists s.(y\ne 0\land u(x,y-1,s)\land r=s+x).
\]
Thus, the satisfiability of \(S\) is equivalent to:
\[
\forall x,y,r.\; \form(x,y,r)\land x> 0 \imp r\ge y,
\]
which is also equivalent to:
\[
\forall x,y,r.\; \dual{\form}(x,y,r)\lor x\le 0 \lor r\ge y,
\]
where \(\dual{\form}\) is the de Morgan dual of \(\form\), given as:
\[
\nu u.\lambda (x,y,r).(y\ne 0\lor r\ne 0) \land
\forall s.(y= 0\lor u(x,y-1,s)\lor r\ne s+x).
\]
In this manner, the satisfiability problem for any CHCs on integer arithmetic
can be converted to the validity checking problem for
a formula of the first-order, \(\nu\)-only fragment (i.e. the fragment without
\(\mu\)) of \HFLZ{};
recall that universal quantifiers can be expressed by using \(\nu\).
Conversely, the validity checking problem for
any formula of the first-order, \(\nu\)-only fragment
where the types of fixpoint variables are restricted to those of the form
\(\INT\to\cdots\to\INT\to\Prop\)
can be reduced to the satisfiability problem for CHCs on integer
arithmetic~\cite{DBLP:conf/sas/0001NIU19}.
In this sense,
the \HFLZ{} validity checking problem can be considered
a generalization of
the CHC satisfiability problem, where higher-order predicates and
fixpoint alternations between \(\mu\) and \(\nu\) are allowed.

Burn et al.~\cite{DBLP:journals/pacmpl/BurnOR18,DBLP:conf/lics/OngW19}
recently studied a higher-order extension of CHCs called HoCHC.
The \HFLZ{} validity checking problem can also be considered
an extension of the satisfiability problem for HoCHC with fixpoint alternations.

\section{From Program Verification to HFL(Z) Model Checking}
\label{sec:red}

This section explains how \HFLZ{} model/validity checking can be applied to
program verification. As seen at the end of the last section,
\HFLZ{} validity checking subsumes CHC solving, which already have a plenty of
applications to program verification~\cite{Bjorner15},
but how can we exploit the additional power of \HFLZ{} model/validity checking
for program verification?

A standard approach to applying model checking to program verification is
to model a program as a transition system, and a property of the program to be
checked as a specification; that is indeed the case for applications of
finite-state model checking~\cite{ClarkeModelChecking},
pushdown model checking~\cite{SLAM}, and
HORS model checking  (which is another kind of higher-order extension
of model checking)~\cite{Ong06LICS,Kobayashi13JACM}.
In applying \HFLZ{} model checking to higher-order program verification,
we actually switch the roles of systems and specifications:
a program is mapped to a \HFLZ{} formula, and a property is mapped to
a finite state system, where the \HFLZ{} formula is a kind of
``characteristic formula'' of the program. This has been partially inspired
by the correspondence between HFL{} model checking and HORS model checking,
where we also need to switch the roles of systems and specifications~\cite{Kobayashi17POPL}.

\newcommand\READ{\mathtt{read}}
\newcommand\CLOSE{\mathtt{close}}
\newcommand\END{\mathtt{end}}

\begin{figure}[tb]
  \begin{center}
    \includegraphics[scale=0.4]{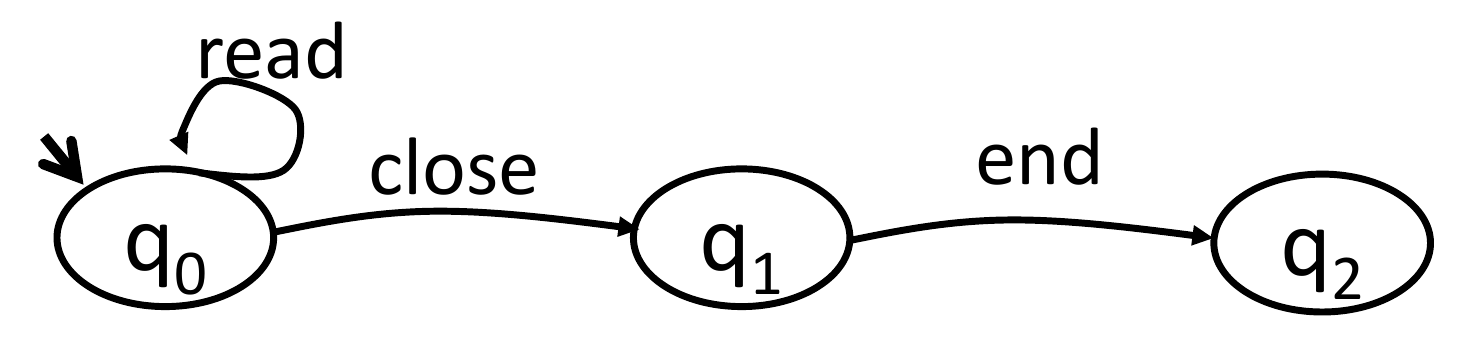}
    \end{center}
\caption{An LTS $\M_{\textit{file}}$ expressing the valid file access protocol}
\label{fig:file}
\end{figure}

Let us consider the following file-accessing program, taken from \cite{KTW18ESOP}.
\begin{verbatim}
  let x = open "foo" in (read(x); read(x); close(x))
\end{verbatim}
This program opens the file ``foo'', and then reads and closes
the file. Suppose we wish to check that the file ``foo'' is indeed
accessed as a read-only file. To this end,
we express the valid access protocol for a read-only file pointer as a labeled
transition system, as shown in Figure~\ref{fig:file}.
The LTS corresponds to a deterministic automaton that accepts the valid traces
\(\READ^*\cdot \CLOSE\cdot \END\).
Here, \texttt{end} denotes the termination of a program.
In the state \(q_0\) (which is the initial state immediately after a file is opened),
 both \texttt{read} and \texttt{close} operations are allowed, but
 after the \texttt{close} operation, only the \texttt{end} operation
 is allowed. 
The program can be converted to the \HFLZ{} formula:
\[ \Some{\texttt{read}}\Some{\texttt{read}}\Some{\texttt{close}}\Some{\texttt{end}}\TRUE,\]
which intuitively
means that the program follows the access protocol represented by the LTS.
It is obtained by just replacing each of the read, close, and end operations
with the corresponding modal operators that say ``those actions are allowed in
the current state.''\footnote{Please ignore the open operation here; it matters
  when more than one file is used in a program, as in \cite{Kobayashi13JACM}.}
It is easy to see that the program accesses the file ``foo'' just if
\[\M_{\textit{file}} \models \Some{\texttt{read}}\Some{\texttt{read}}\Some{\texttt{close}}\Some{\texttt{end}}\TRUE.\]

The same idea applies to more complex programs that contain recursion
and conditionals.
Let us consider the following program:
\begin{verbatim}
  let x = open "foo" in 
  let rec f n = if n<=0 then close x else (read x; f (n-1) x) in
  f 10 
\end{verbatim}
To make the reduction clearer, let us write it in the continuation passing style.
\begin{verbatim}
  let x = open "foo" in 
  let rec f n k = if n<=0 then close x k else read x (f (n-1) x k) in
  f 10 ()
\end{verbatim}
Here, \texttt{read} and \texttt{close} now take an additional continuation parameter,
which is invoked after the read/close operations.
Then, the property that the program accesses the file ``foo'' as a read-only file
just if:
\[
\M_{\textit{file}}\models
(\mu f.\lambda n.\lambda k.(n\le 0 \To \Some{\CLOSE}k)\land (n>0\To \Some{\READ}(f\,(n-1)\,k)))
10\, (\Some{\END}\TRUE).\]
Here, the formula contains integers and order-1 fixpoint operators.
As before, the formula has been obtained by just replacing
 each of the read/close operations, and program
termination (represented by \texttt{()}) with the corresponding modal operator.
The conditional ``\texttt{if n<=0 then ... else ...}'' has been replaced
by the corresponding logical formula \((n\le 0 \To \cdots)\land (n>0\To \cdots)\),
and the recursion has been replaced by the fixpoint operator \(\mu\) (here,
by using \(\mu\), we require that the program terminates).

We have given above just order-1 examples, but it should be clear that the idea
of the translation should work for higher-order programs. A general translation
for linear-time properties is found in \cite{KTW18ESOP},
a translation for arbitrary $\omega$-regular properties (including both
linear-time and branching time properties)
is found in \cite{DBLP:conf/pepm/WatanabeTO019}.
For linear-time properties of first-order recursive programs,
a more optimized translation is given in \cite{DBLP:conf/sas/0001NIU19}.
In those general translations, program verification problems are
actually reduced to the \emph{validity} checking problem for \HFLZ{} formulas,
by using a kind of product construction.

\section{Solving HFL(Z) Model Checking Problems}
\label{sec:solve}

In this section, we discuss how to solve instances of the \HFLZ{} model checking
problem obtained from program verification problems.
For the sake of simplicity, we actually focus on the  validity checking problem
(which is a special case of the model checking problem where the formula contains
no modal operators; recall Section~\ref{subsec:HFL}),
but most of the techniques apply to the model checking problem as well.
Some of our tools and benchmark sets mentioned below are
available from \url{https://github.com/hopv}. For the other tools,
please consult each paper cited below.

\begin{figure}[tb]
  \begin{center}
    \includegraphics[scale=0.5]{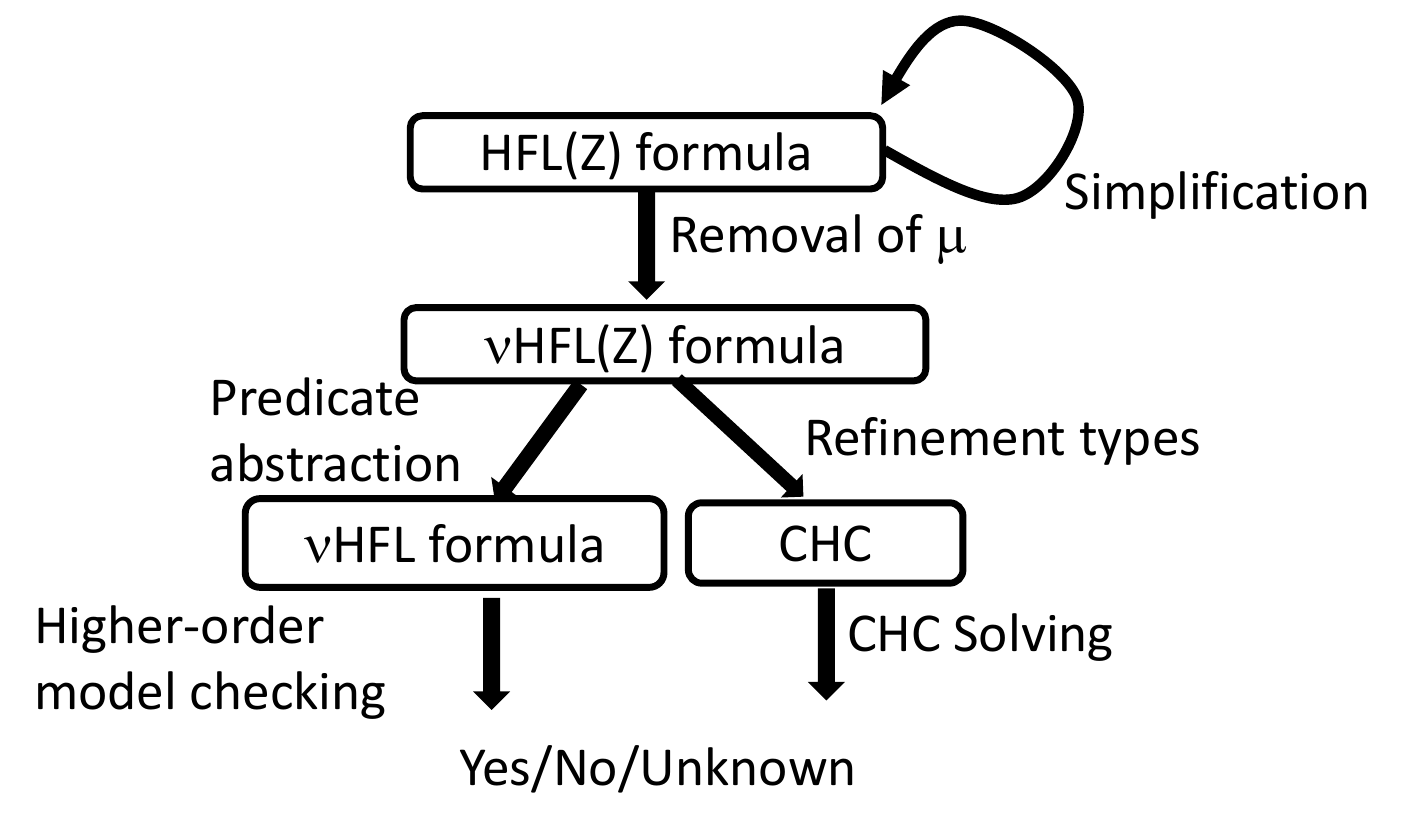}
    \end{center}
\caption{An overview of our method for \HFLZ{} validity checking}
\label{fig:validitychecking}
\end{figure}

Our overall method for \HFLZ{} validity checking is summarized in
Figure~\ref{fig:validitychecking}. In the figure,
a ``$\nu$\HFLZ{} formula'' refers to a \HFLZ{} formula without the least
fixpoint operator \(\mu\). 
The overall strategy for solving the validity checking problem is
analogous to, and has been inspired by automated program verification methods.
The first phase of removing the least fixpoint operator \(\mu\) corresponds
to reductions from liveness property verification (such as termination verification)
to safety property
verification~\cite{DBLP:conf/lics/PodelskiR04,Cook07POPL,Kuwahara2014Termination}.
The two methods for checking the validity of \nuHFLZ{} formula
correspond to two major approaches to automated verification of higher-order
programs: higher-order model checking~\cite{KSU11PLDI} and refinement types~\cite{Hongwei99POPL,Unno09PPDP,Jhala08,DBLP:conf/popl/UnnoTK13}.
We discuss each step of Figure~\ref{fig:validitychecking} below.

Due to underapproximation in various steps
(a formula \(\form\) is replaced by another formula \(\form'\) such that
\(\form'\To\form\), so that the validity of \(\form'\) implies that
of \(\form\), but not vice versa),
the procedure shown in Figure~\ref{fig:validitychecking}
 cannot conclude that the original formula \(\form\) is invalid
even if an approximation of the formula is invalid.
Thus, given a \HFLZ{} formula \(\form\), we also prepare its de Morgan dual
\(\dual{\form}\), and apply the whole procedure in parallel to \(\form\)
and \(\dual{\form}\).
If \(\dual{\form}\) is valid, then we can conclude that \(\form\) is invalid.

\subsection{Removing $\mu$}
In the reductions from program verification problems to \HFLZ{} validity
checking~\cite{KTW18ESOP,DBLP:conf/pepm/WatanabeTO019,DBLP:conf/sas/0001NIU19},
liveness and safety properties are respectively turned into \(\mu\)- and \(\nu\)-formulas.
Thus, following the techniques for liveness property verification~\cite{DBLP:conf/lics/PodelskiR04,Cook07POPL,Kuwahara2014Termination,DBLP:conf/popl/MuraseT0SU16,freqterm},
it is natural to first remove \(\mu\)-formulas by using analogous techniques.

In \cite{DBLP:conf/sas/0001NIU19}, we have adopted the technique
of Fedyukovich et al.~\cite{freqterm} for the first-order fragment of \HFLZ{} formulas.
Suppose that we wish to prove the validity of
a \(\mu\)-formula of the form \(\mu x.\form(x)\).
By the standard fixpoint theorem, we have
\[\form^{n}(\FALSE) \To \mu x.\form(x)\]
for any natural number \(n\).
The formula \(\form^{n}(\FALSE)\) is equivalent to
\((\nu x'.\lambda z.(z>0\land \form(x'(z-1))))n\),
which is also equivalent to the following formula \(\psi\):
\[\psi \DEF \forall u\ge n.(\nu x'.\lambda i.(i>0\land \form(x'(i-1))))u.\]
Thus, it suffices to show that \(\psi\) is valid.
Here, \(n\) can be considered a bound for the number of unfoldings of the
original \(\mu\)-formula; by gradually increasing \(n\), we can obtain
a better approximation of the original formula.

In the case of the first-order \HFLZ{} formulas, the above translation
yields order-1 \nuHFLZ{} formulas (where all the types are of the form
\(\INT^k\to \Prop\)), whose validity checking problems can be further
reduced to the CHC satisfiability problem, as discussed in Section~\ref{sec:hfl}.
\begin{example}
  Let us consider proving the validity of the formula:
  \[ \forall i.(\mu x.\lambda y.y\le 0\lor x(y-1))i.\]
  Based on the above translation, it suffices to show:
  \[ \forall i.\forall u\ge \max(i+1,1).
    (\nu x'.\lambda (z,y).z>0 \land (y\le 0\lor x'(z-1,y-1)))(u,i).\]
  Here, \(\max(i+1,1)\) corresponds to the bound \(n\) above;
  in general, the bound may depend on free variables as in this example.
  Note that the resulting formula contains only \(\nu\) and \(\forall\),
  where the latter can also be expressed by \(\nu\),
  as seen in Example~\ref{ex:quantifiers}.

  To translate the formula above to CHCs, observe that the formula is equivalent to:
  \[ \forall i.\forall u\ge \max(i+1,1).
  (\mu \dual{x'}.\lambda (z,y).z\le 0 \lor (y> 0\land \dual{x'}(z-1,y-1)))(u,i)
  \To \FALSE.\]
  The \(\mu\)-formula
\(\mu \dual{x'}.\lambda (z,y).z\le 0 \lor (y> 0\land \dual{x'}(z-1,y-1)))\)
is the least predicate \(X\) that satisfies the following clauses:
\[
\begin{array}{rcl}
  z\le 0 &\imp& X(z,y)\\
  y>0\land X(z-1,y-1) &\imp& X(z,y).
  \end{array}
\]
Thus, the validity of the formula above is
equivalent to the satisfiability of the two clauses above
with the goal clause:
\[
\begin{array}{rcl}
  u\ge i+1\land u\ge 1 \land X(u,i) &\imp& \FALSE.
  \end{array}
\]
The above system of CHCs is indeed satisfiable, and has the model:
\(X(z,y) \DEF z\le 0 \land z\le y\). We can thus conclude that the original
\HFLZ{} formula is valid. \qed
\end{example}

We have implemented the above method for the first-order fragment of
\HFLZ{} and applied it to automated verification of temporal
properties~\cite{DBLP:conf/sas/0001NIU19}. Despite the generality of
the approach (which works for arbitrary \(\omega\)-regular properties
of while-programs), our implementation generally outperformed
Cook and Koskinen's method specialized for CTL verification~\cite{Cook2013a},
probably thanks to the recent advance of CHC solvers~\cite{DBLP:journals/fmsd/KomuravelliGC16,DBLP:journals/jar/ChampionCKS20,Eldarica}
and the streamlined approach.
Work is under way to extend the translation above for \HFLZ{} formulas
of arbitrary orders.
\subsection{Predicate Abstraction for \nuHFLZ{} Validity Checking}

One approach~\cite{DBLP:conf/sas/IwayamaKST20} to proving the validity of a \nuHFLZ{} formula \(\form\) is
to apply predicate abstraction to obtain a pure \nuHFL{} formula \(\form'\)
(i.e., a \nuHFLZ{}
formula without integers) as an underapproximation of \(\form\),
and then apply an algorithm for pure HFL model
checking~\cite{DBLP:conf/aplas/Hosoi0T19}\footnote{In the actual implementation,
  we actually use a HORS model checker~\cite{horsat2,Kobayashi13horsat}
  based on the correspondence between HFL and HORS model checking~\cite{Kobayashi17POPL}.}
(recall that pure HFL  model checking is decidable; despite its high worst-case
complexity, practical algorithms exist, which do not always suffer from the high complexity).
This approach may be viewed as a generalization of
the HORS model checking approach to (un)reachability verification~\cite{KSU11PLDI}
and non-termination verification~\cite{Kuwahara2015Nonterm}.

Given a set of predicates on integers, a given \nuHFLZ{} formula can be
underapproximated by a pure HFL formula.
For example, suppose that we have decided to abstract
every integer with  the predicate \(\lambda y.y> 0\).
Then, the formula 
\(\phi \DEF (\nu x.\lambda y.y\ge 0 \land x(y+1))1\)
can be underapproximated by \(\phi'\):
\[(\nu x.\lambda b. b\land x(b))\TRUE,\]
where \(b\) is a Boolean variable corresponding to the condition \(y>0\).
Since \(\phi'\) is valid (as can be confirmed by a pure HFL model checker),
we can conclude that the original formula \(\phi\) is also valid.
As in standard approaches to combining predicate abstraction and model checking,
predicates to be used for abstraction can be found in a counterexample-guided manner.
More details can be found in \cite{DBLP:conf/sas/IwayamaKST20}.

\subsection{Refinement Types for \nuHFLZ{} Validity Checking}

We have also studied another approach to \nuHFLZ{} validity checking,
based on a (sound but incomplete) reduction to a refinement type inference
problem~\cite{DBLP:conf/aplas/KatsuraIKT20}.
The approach has been inspired by the refinement type system of
Burn et al.~\cite{DBLP:journals/pacmpl/BurnOR18} for HoCHC,
whose idea can further be traced back to refinement type systems for
functional programs~\cite{Hongwei99POPL,Unno09PPDP,Jhala08,DBLP:conf/popl/UnnoTK13}.

The syntax of refinement types for \nuHFLZ{} is given by:
\[
\sigma ::= \Prop[\psi] \mid \sigma_1\to \sigma_2 \mid x\COL\INT\to \sigma.
\]
Here, \(\psi\) is a quantifier-free formula of integer arithmetic (which
may contain integer variables bound by \(x\COL\INT\to\cdots\)).
The type \(\Prop[\psi]\) describes propositions that hold whenever
\(\psi\) holds. For example, the \nuHFL{} formula
\(x\ge 0\) has type \(\Prop[x>0]\), since \(x\ge 0\) holds whenever
\(x>0\) holds. A predicate \(\lambda p.p(0)\) has type
\((x\COL\INT\to \Prop[x\ge 0])\to \Prop[\TRUE]\), because
\(p(0)\) holds whenever \(p\) is a predicate on integers such that
\(p(x)\) holds for every \(x\ge 0\).
Based on the intuition, one can construct a refinement type system for
\nuHFLZ{}, such that a \nuHFLZ{} formula \(\form\) is valid
if \(\form\) has type \(\Prop[\psi]\), and reduce the type inference problem 
 to a constraint satisfaction problem on predicate variables
in a standard
manner~\cite{Unno09PPDP,DBLP:journals/pacmpl/BurnOR18,DBLP:journals/jar/ChampionCKS20}.
Unlike the case of refinement type systems for
(un)reachability verification of functional
programs~\cite{Unno09PPDP,DBLP:journals/jar/ChampionCKS20},
the resulting constraint satisfaction problem is no longer a CHC problem in general;
constraints on predicates may be of the form:
\[
P_1(\seq{x},\seq{y})\land \cdots \land P_k(\seq{x},\seq{y})
\land \psi(\seq{x},\seq{y}) \imp Q_1(\seq{x})\lor \cdots \lor Q_\ell(\seq{x}),
\]
where disjunction may occur in the head
(here, \(P_i, Q_j\) are unknown predicate variables
and \(\psi(\seq{x},\seq{y})\) is a formula of integer arithmetic).
Solving this generalized form of constrained clauses~\cite{DBLP:conf/aaai/Satake0Y20}
is the current major bottleneck
of this approach; work is under way to extend the ICE-based CHC solving
approach~\cite{DBLP:journals/jar/ChampionCKS20,DBLP:journals/pacmpl/EzudheenND0M18} to deal
with the generalized constrained clauses.

Despite the bottleneck mentioned above, the refinement type-based approach
to \nuHFLZ{} validity checking is generally faster than the predicate
abstraction-based approach, while the latter tends to be more precise.
Thus, the two approaches are complementary to each other.

\subsection{Unfold/fold Transformations for Simplification}

Inspired by the unfold/fold transformation techniques
for CHC solving~\cite{DBLP:journals/tplp/AngelisFPP18a,DBLP:conf/cade/AngelisFPP20},
we have also studied unfold/fold transformations for the first-order fragment
of \HFLZ{}~\cite{DBLP:conf/tacas/KobayashiFG20} to enhance the power
of an automated \HFLZ{} validity checker.
The transformations are useful for reasoning about relations between
fixpoint formulas.

\newcommand\Even{\mathit{Even}}
\newcommand\DEven{\dual{\Even}}
\newcommand\Odd{\mathit{Odd}}
For example, consider proving \(\forall n.\Even(n) \To \Odd(n+1)\), where
\(\Even\) and \(\Odd\) are defined by:
\[
\begin{array}{rcl}
  \Even &\DEF& \mu x.\lambda y.(y=0\lor x(y-2))\\
  \Odd & \DEF& \mu x.\lambda y.(y=1\lor x(y-2)).
\end{array}
\]
It can be expressed as the \HFLZ{} formula \(\forall n.\DEven(n)\lor \Odd(n+1)\),
where
\[
\begin{array}{rcl}
  \DEven &\DEF& \nu x.\lambda y.(y\ne 0\land x(y-2)).
  \end{array}
  \]
  Let \(\DEven(y)\lor \Odd(y+1)\) be \(\form(y)\). It can be transformed as follows:
\[
\begin{array}{lcll}
  \form(y) &\equiv&
  \DEven(y)\lor \Odd(y+1)\\
  &\equiv& (y\ne 0\land \DEven(y-2))\lor (y+1=1\lor \Odd(y-1)) & \mbox{(unfold
    $\DEven$ and $Odd$)}\\
  &\equiv& y+1=1\lor \DEven(y-2)\lor \Odd(y-1) & \mbox{(shuffle the formula)}\\
  &\equiv& y=0 \lor \form(y-2) & \mbox{(by the definition of $\form$).}
\end{array}
\]
Based on the transformations above, we can replace 
\(\DEven(n)\lor \Odd(n+1)\) with
\((\nu x.\lambda y.y=0\lor x(y-2))n\), which is obviously valid
(since \(\nu x.\lambda y.y=0\lor x(y-2)\equiv \lambda y.\TRUE\)).
The above sequence of transformations are analogous to unfold/fold
transformations for CHCs~\cite{DBLP:journals/tplp/AngelisFPP18a},
but the soundness of the overall transformations is more subtle,
due to the mixture of the least and greatest fixpoint operators:
see \cite{DBLP:conf/tacas/KobayashiFG20} for the conditions
of soundness of unfold/fold transformations for the first-order
fragment of \HFLZ{}.

\subsection{Semi-Automated Methods}
We have so far discussed automated methods for \HFLZ{} validity checking.
As the \HFLZ{} formula obtained from a program verification problem can
be considered a kind of ``verification condition,''\footnote{Thanks to
  fixpoint operators, no annotations of loop invariants and pre/post-conditions
  of recursive functions are required.}
it is also natural to prove the validity of the formula semi-automatically,
possibly using a proof assistant such as Coq,
as exploited in our recent
work~\cite{DBLP:conf/pepm/WatanabeTO019,DBLP:conf/csl/KoriT021}. 
Integration with the automated methods is left for future work.

\section{Conclusion}
We have given an overview of our project on automated program verification
based on \HFLZ{} model and validity checking. Our framework can be considered
a generalization of the CHC-based program verification framework
and provides a uniform approach to higher-order program verification.
One may wonder whether \HFLZ{} is too expressive as the target of reductions
from program verification problems. To answer the question,
Tsukada~\cite{DBLP:conf/lics/Tsukada20}
has recently shown that, in a certain sense, \HFLZ{} is just as expressive as needed
for encoding higher-order program verification problems.
A lot of work is still left to be done, including
a full implementation of the \HFLZ{} validity checker and 
further improvement of
backend solvers for CHCs and generalized constrained clauses.

\section*{Acknowledgment}
We would like to thank anonymous referees for useful comments.
This work was supported by JSPS KAKENHI Grant Number 
JP20H00577 and JP20H05703.


\end{document}